\documentclass{pnastwo}

\usepackage[dvips]{graphicx}
\usepackage{PNAStwoF}

\usepackage[varg]{txfonts}

\newcommand{\vk}{{\bf k} }

\begin{document}

\conflictofinterest{Conflict of interest footnote placeholder}

\track{This paper was submitted directly to the PNAS office.}

\footcomment{Abbreviations: 2D, two dimensional;
MOSFET, metal oxide semiconductor
field effect transistor; RPA, random phase approximation}

\title{A self-consistent theory for graphene transport}

\author{Shaffique Adam\affil{1}{Condensed Matter Theory 
Center, Department of Physics, University of Maryland,
College Park, MD 20742-4111, USA}\thanks{To whom correspondence 
should be addressed. E-mail: adam1@umd.edu},
E. H. Hwang\affil{1}{}, V. M. Galitski\affil{1}, \and
S. Das Sarma\affil{1}{}}

\contributor{Submitted to Proceedings of the National Academy of Sciences
of the United States of America}

\maketitle

\begin{article}

\begin{abstract}
We demonstrate theoretically that most of the observed transport
properties of graphene sheets at zero magnetic field can be explained
by scattering from charged impurities.  We find that, contrary to common
perception, these properties are not universal but depend on the
concentration of charged impurities $n_{\rm imp}$.  For dirty samples
($250~\times 10^{10}~{\rm cm}^{-2} < n_{\rm imp} < 400~\times
10^{10}~{\rm cm}^{-2}$), the value of the minimum conductivity at low
carrier density is indeed $4 e^2/h$ in agreement with early
experiments, with weak dependence on impurity concentration.  For
cleaner samples, we predict that the minimum conductivity depends
strongly on $n_{\rm imp}$, increasing to $8 e^2/h$ for $n_{\rm imp}
\sim 20~\times 10^{10}{\rm cm}^{-2}$.  A clear strategy to improve
graphene mobility is to eliminate charged impurities or use a
substrate with a larger dielectric constant.  
\end{abstract}


\dropcap{T}
he past two years have seen a proliferation of theoretical and
experimental interest in graphene.  The interest stems mainly from the
striking differences between graphene and other more well known
semiconductor-based two dimensional (2D) systems, that arise mostly
from its unique band structure, obtained by considering graphene to be
a single sheet of carbon atoms arranged in a honey-comb lattice.
Graphene is in fact a carbon nanotube rolled out into a single 2D
sheet, and as was already known from studying carbon nanotubes,
electrons moving in the periodic potential generated by the carbon
lattice form a band that displays striking properties such as having a
chiral Dirac equation of motion with a mathematical structure similar
to Weyl neutrinos.  This intriguing `{\it relativistic}' Dirac-Weyl
spectrum of graphene has attracted substantial interest and attention.
While this analogy of considering graphene as a solid-state
realization of the ``{\it massless chiral Dirac Fermion}'' model
(developed as a solution to Dirac's Lorentz invariant generalization
of Schr\"{o}dinger's equation) has some utility, it also has the
potential to be misleading.  In particular, as we argue here,
searching to explain the experimental transport properties of graphene
by focusing on the ``{\it Dirac point}'' (see formal definition
below), obscures the real mechanism of carrier transport.  In this
respect, and from our perspective, the physics of graphene has more in
common with the Metal-Oxide-Semiconductor-Field-Effect-Transistors
(MOSFETs) that form the backbone of our current day semiconductor
industry, than with the physics of relativistic chiral Fermions.
Studying graphene is therefore as much about making useful MOSFETs from
pencil smudges, as it is about studying quantum electrodynamics 
in a pencil mark.

We observe that already within one year since the fabrication of the
first gated 2D graphene samples (that enable a variable external gate
voltage tuned carrier density), mobilities as high as $2.5~{\rm
m^2}/{\rm Vs}$ have been reported, and these values are comparable to
the best Si MOSFET samples at low temperature.  In addition, graphene
mobility is relatively temperature independent, making
room temperature 2D graphene mobilities to be among the highest in FET
type devices.  It is therefore both of fundamental and technological
interest to understand the transport mechanism in graphene in
reasonable qualitative and quantitative detail.  Similar to MOSFETs,
transport properties of graphene are determined by scattering from
charged impurities which are invariably present.  We report here the
essential graphene transport theory focusing on charged impurity
scattering.

Prior to this work, the conventional wisdom in the graphene community
was that close to the Dirac point, carrier transport is ballistic,
that the minimum conductivity is universal and that we lack a basic
understanding of how at the Dirac point there could be carrier free
transport over micron-sized distances.  In this context, our work
provides a theoretically simple explanation for this graphene
transport mystery: charged impurities in the substrate generate
carrier density fluctuations that allow for non-universal diffusive
transport, and that these density inhomogeneities render the Dirac
point physics experimentally inaccessible, at least for current
graphene samples (see note added at the end of this paper).

We emphasize that although the importance of charged impurity
scattering in determining the linear-in-density high-density (i.e.
far from the charge neutral Dirac point) graphene carrier transport
has already been
established~\cite{kn:ando2006,kn:nomura2007,kn:cheianov2006,kn:hwang2006c}
in the literature, the current work is the first to point out that the
same charged impurities will have a qualitative effect at low carrier
density close to the Dirac point by providing an inhomogeneous
electron-hole puddle landscape where the conductivity will be
approximately a constant over a finite range of external gate voltage,
providing a simple and physically appealing explanation for the
observed graphene minimum conductivity plateau.  The importance of our
work therefore lies in its ability to explain the graphene transport
data at low carrier density through a physically appealing charged
impurity-induced mechanism which quantitatively explains both the
existence (and the width) of the minimum conductivity plateau as well
as its magnitude.  No such explanation existed in the literature
before our work, and the minimum conductivity phenomenon in graphene
was considered to be an outstanding experimental puzzle.

\begin{figure}[t]
\vspace{0.05in}
\begin{center}
\centerline{\includegraphics[width=.4\textwidth]{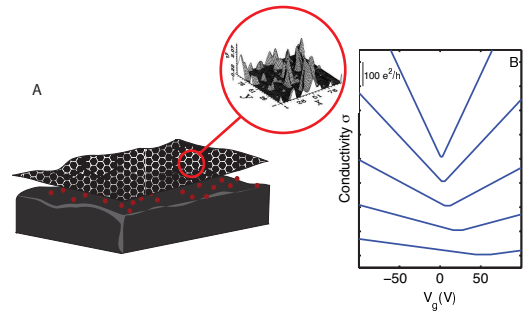}}
\caption{\label{Fig:model} 
(A) Main panel shows cartoon of
our model where charged impurities in the substrate at a distance $d$
away from the graphene sheet create a spatially inhomogeneous screened
Coulomb potential.  At low carrier density, the system breaks up into
puddles of electrons and holes.  The residual density is then calculated
self-consistently giving remarkable agreement with experimental
results.  The inset (taken from~\protect{\cite{kn:efros1993}}) is used
here just to illustrate the voltage fluctuations schematically, where
it is understood that the presence of both electron and hole carriers
implies that both positive and negative voltages are screened. (B)
Predicted conductivity traces (Eq.~\protect{\ref{Eq:Main}}) for
different values of $n_{\rm imp}$.  Curves are offset 
vertically by $100 e^2/h$ for clarity and show from top to bottom,
impurity concentrations (in units of $10^{10}~\rm{cm}^{-2}$):
20 (very clean), 40, 80, 160, 320 (very dirty).  Most of the samples
considered in Ref.~\protect{\cite{kn:novoselov2005}} had mobilities
in between the bottom two curves.  Clearly seen are the gate voltage
offset $V_g^D$ and the minimum conductivity plateau, which are 
larger for the dirtier samples.}
\end{center}
\end{figure}

In this context, ``{\it bare graphene}'' is the empty honeycomb
 lattice, where allowing for electron hoping between adjacent sites
 gives the linear Dirac-Weyl spectrum. The nominally undoped or
 ungated situation is that of the completely filled valence band and a
 completely empty conduction band which touch at the Dirac point
 making graphene a zero-gap semiconductor.  The Dirac point is a
 singular point of measure zero that separates the conduction and the
 valence band in the linear graphene spectrum.  This ``{\it intrinsic
 graphene}'' with the chemical potential (or Fermi energy) precisely
 at the Dirac point has no free carriers and is obviously an abstract
 model, since the slightest amount of doping or external potential
 will induce carriers in the system.  Charged impurity disorder or
 spatial inhomogeneity will render this intrinsic graphene
 experimentally unrealizable.  ``{\it Extrinsic graphene}'' is when
 one induces free carriers (either electrons in the conduction band
 with a positive potential, or holes in the valence band with a
 negative potential), by applying an external gate voltage $V_g$, or
 equivalently, by doping the system.  Carrier bands are filled up to a
 certain Fermi Energy $E_{\rm F}$, determined by the electrostatic
 potential configuration of the graphene environment, and depending on
 which is larger, could be dominated either by the external gate
 voltage or by charged impurities.  All experimental graphene samples
 are extrinsic, since there are invariably some free carriers present
 in the system, and transport close to the Dirac point is dominated by
 two distinct effects of the charged impurities in the system: (i) the
 induced graphene carrier density is self-consistently determined by
 the screened charged impurity potential; and (ii) the conductivity is
 determined by charged impurity scattering.

As one would expect with any newly discovered electronic
material~\cite{kn:novoselov2004}, the full gamut of experimental
techniques has been used to explore graphene properties, including light
scattering~\cite{kn:yan2006}, angle resolved photo-emission
spectroscopy~\cite{kn:bostwick2007}, and surface probe
measurements~\cite{kn:ishigami2007,kn:martin2007}. The field continues
to evolve with new ideas being explored including making suspended
graphene~\cite{kn:meyer2007} and electromechanical
resonators~\cite{kn:bunch2007}, transfer-printing graphene onto
plastic~\cite{kn:chen2007}, using
superconducting~\cite{kn:heersche2007} and 
ferromagnetic leads~\cite{kn:hill2006}, having
patterned top-gates~\cite{kn:lemme2007,kn:huard2007,kn:williams2007}, 
exposing samples to molecular
dopants~\cite{kn:schedin2006} and fabrication of
graphene-nano-ribbons~\cite{kn:han2007}.  These rapid experimental
advances show that the study of graphene is still in its infancy, with
many promises for the discovery of new physics and for application to
technology.

The focus of the present work is on the important graphene transport
measurements~\cite{kn:novoselov2005, kn:zhang2005}.  These were the
first experiments to be done, and are still not understood, in that
there are widely disparate claims on the transport mechanism.  Since
these experiments form the basis for most of the future work on
graphene as well as its prospective technological applications, a
correct understanding of the basic transport physics is of fundamental
importance.  While there are many features observed in the transport
experiments, two have been highlighted and particularly discussed in
the literature, namely, the low-density ``{\it minimum conductivity}''
$\sigma_0$, i.e. the value of the conductivity at or near the Dirac
point (where $E_{\rm F} \sim 0$, and a na\"{i}ve picture would suggest
that there are no charge carriers); and the high-density conductivity
$\sigma(n)$ which is linear in the carrier density $n$, giving a
constant mobility $\mu = \sigma/ne$.  As we discuss below, our
analytic theory is the first to explain both features quantitatively,
as well as make predictions for the width of the minimum conductivity
plateau and the offset of the Dirac point from zero gate voltage.

In the literature, most theoretical work has focused on the
short-range scattering mechanism (also called ``{\it white noise}''
disorder) to understand graphene transport, mainly as a matter of
technical convenience.  Early work using a Kubo formalism in the
ballistic limit~\cite{kn:fradkin1986,kn:ludwig1994} showed that the
conductivity for massless Dirac Fermions is $e^2/(\pi h)$ for
vanishing disorder, and that this universal value occurs only at the
Dirac point and not in its vicinity.  At finite carrier density, the
Kubo formalism with short-range scattering gives a conductivity that
is constant with carrier density~\cite{kn:ando2006} and not the linear
in density behavior seen in experiments.  Certain
numerical~\cite{kn:nomura2007} and analytical~\cite{kn:ziegler2006}
methods that try to extrapolate between these two limits inevitably
get a square-root dependence of conductivity on density (not
linear), and give orders-of-magnitude incorrect values for the
mobility.  More recently, short range scattering has been considered
theoretically~\cite{kn:aleiner2006,kn:altland2006,kn:ostrovsky2006}
with the finding that at zero temperature, localization effects should
give $\sigma_0 = 0$.  While all these works improve our abstract
theoretical understanding of graphene, they are all in qualitative
disagreement with existing experimental data.  We argue here that
short-range scattering has little to do with the experiments of
Refs.~\cite{kn:novoselov2004,kn:novoselov2005,kn:zhang2005}, and that
although localization effects may very well be important in the zero
temperature limit, the existing bulk graphene transport data at
accessible temperatures ($T > 0.1 K$) are in the Drude-Boltzmann
diffusive transport regime.  Equally important, the observed transport
properties of doped graphene do not access the Dirac point physics, at
least in the currently available
samples~\cite{kn:novoselov2005,kn:zhang2005} which
have fairly large concentrations of charged impurity centers.  While
the observed conductivity value (and not conductance) of $4e^2/h$ in
dirty samples brings to mind connections with universal 
conductance quantization phenomena such as one-dimensional
point-contact conductance quantization or Zitterbewegung or
quantized Hall resistance, we argue here that no such universal
physics is at play in current bulk graphene transport experiments where
conductivity, and not conductance is being discussed.    

Our goal here is to develop a quantitatively accurate analytic
theory for the most important graphene transport problem, namely, the
regular bulk dc conductivity studied in the diffusive Drude-Boltzmann
limit, which is the regime of technological interest.  We argue that
charged impurity scattering is responsible for most of the observed
bulk diffusive transport behavior in graphene.  Charged impurities
could reside either inside the substrate or created near the
graphene-substrate interface during the processing and handling of
samples.  The typical concentration of charged impurities in a SiO$_2$
substrate is $n_{\rm imp} \sim 50~\times 10^{10} \rm{cm}^{-2}$, and
are known to dominate the transport properties of other extensively
studied 2D semiconductor systems~\cite{kn:ando1982}.

Earlier work~\cite{kn:nomura2007,kn:cheianov2006,kn:hwang2006c}
demonstrated that while the mean free path for short-range scatterers
$\ell_{s} \sim 1/\sqrt{n}$, where $n$ is the carrier density, for
Coulomb (i.e. charged impurity) scatterers $\ell_{c} \sim \sqrt{n}$.
This implies that although graphene at low density is a clean or
ballistic system for short-range scatterers, it is a dirty and
diffusive system for long-range scatterers.  At low carrier densities
of $n \sim n_{\rm imp} \sim 50~\times 10^{10} \rm{cm}^{-2}$, one can
estimate that $\ell_{s} \gtrsim 1000 \ {\rm nm}$ and $\ell_{c} \lesssim
50  \ {\rm nm}$.  Therefore, at the lowest densities close to the Dirac
point, graphene transport properties are completely dominated by
Coulomb scattering.  This simple argument also establishes that any
short-range scattering, even if it is present in graphene, is
irrelevant in the low carrier density limit near the Dirac point where
the minimum conductivity plateau is observed.

In the present work we provide a complete picture of transport in both
high and low carrier density regimes using a self-consistent
RPA-Boltzmann formalism where the impurity scattering by the charged
carriers themselves is treated self-consistently in the Random Phase
Approximation (RPA), and the dc conductivity is calculated in the
Boltzmann kinetic theory.  We derive analytic expressions for (i)
mobility $\mu$, (ii) plateau width, (iii) minimum conductivity, and
(iv) shift in gate voltage.  Our results for graphene on a SiO$_2$
substrate can be summarized as

\begin{eqnarray}
\label{Eq:Main}
\sigma(n - \bar{n}) = \left\{ \begin{array}{l}
     \frac{20 e^2}{h} \frac{n^*}{n_{\rm imp}} 
           \ \ \  \mbox{if $n - \bar{n}< n^*$}, \\
       \frac{20 e^2}{h} \frac{n}{n_{\rm imp}} 
            \ \ \ \mbox{if $n - \bar{n} > n^*$},
\end{array} \right.
\end{eqnarray}
where Eq.~\ref{Eq:SC} below gives analytic expressions for $n^*$ and
$\bar{n}$.

An important finding of our work is that for dirty samples with
$n_{\rm imp} \sim 3.5~\times 10^{12} \rm{cm}^{-2}$, $\sigma_0$ is
indeed close to $4 e^2/h$ as observed experimentally, and not very
sensitive to changes in disorder, while for cleaner samples with
$n_{\rm imp} \sim 2~\times 10^{11} \rm{cm}^{-2}$, $\sigma_0 \sim 8
e^2/h$, and is sensitive to the value of $n_{\rm imp}$, thus
explaining the mystery of why more recent experiments show a larger
magnitude and larger spread in the value of $\sigma_0$.  For the
typical densities used in the early graphene experiments, the minimum
conductivity appears to saturate at a universal value of $4 e^2/h$,
but we predict that there is nothing universal here, and for dirtier
samples, the value of the minimum conductivity as a function of
$n_{\rm imp}$ would slowly decrease.

The formalism developed here to obtain Eq.~\ref{Eq:Main} can be
divided into three steps.  First, we develop an analytical solution
for the Boltzmann transport theory using the full RPA treatment of the
charged impurity scattering.  We find analytically that $\sigma
\approx 20 (e^2/h) (n/n_{\rm imp})$ in agreement with earlier
numerical
calculations~\cite{kn:nomura2007,kn:cheianov2006,kn:hwang2006c}.
Second, we extend the methods of
Refs.~\cite{kn:efros1993,kn:galitski2007} to evaluate the screened
voltage fluctuations induced by charged impurities.  Here we calculate
the potential fluctuations using the full RPA screening, which
although being complicated and cumbersome, is necessary to obtain
quantitative agreement.  Third, we develop a theory to calculate the
residual carrier density $n^*$ self-consistently.  We find that the
ratio $n^*/n_{\rm imp}$, that is directly related to the minimum
conductivity through $\sigma_0 \approx 20 (e^2/h) (n^*/n_{\rm imp})$,
is a monotonically decreasing function of $n_{\rm imp}$ and the
dependence gets weaker for larger impurity density.

Before we provide details of our calculation, we first address the
range of validity of our self-consistent RPA-Boltzmann theory.  First,
we consider only Coulomb scattering.  As already discussed above (see
also Ref.~\cite{kn:hwang2006c}), other scattering mechanisms are
irrelevant at low density and the experimentally observed linear
dependence of conductivity with carrier density singles out charged
impurities as the dominant scattering mechanism.  Only in the limit of
very small charged impurity density, must one include short-range
scattering into the formalism (such short-range scattering may arise
from point-defects and dislocations in the lattice).  We find
$\sigma_0 = (4 e^2/h)[ n_{\rm imp}/ (5 n^*) + \eta]^{-1}$, where we
estimate $\eta = 2/k_{\rm F} \ell_{s} \lesssim 1/10$, suggesting that for
very low impurity densities (two orders of magnitude lower $n_{\rm
imp}$ than present day samples), $\sigma_0$ will saturate at around
$20 e^2/h$, before our charged impurity model gives way to short-range
scattering.  Second, our theoretical calculation is done at zero
temperature. Theoretically one expects~\cite{kn:cheianov2006} very
weak temperature dependence for $T \ll T_{\rm F}$, and this is indeed
consistent with experimental observations where $T/T_{\rm F} \lesssim
0.2$.  Third, since the sample sizes are several microns in length,
and the mean free path is tens of nanometers, we are certainly in the
diffusive as opposed to ballistic transport regime.  Fourth, electron
interactions are treated within the RPA approximation scheme.  RPA is
an expansion in $r_s$, and in graphene experiments on SiO$_2$, $r_s
\approx 0.8$, so one would expect it to work better than for 2D
semiconductors or metals, where although $r_s \sim 2-10$, RPA provides
an excellent approximation.  Fifth, the scattering time is calculated
using Boltzmann kinetic theory.  Formally, Boltzmann theory (as
described by Eq.~\ref{Eq:TI} below) is valid for $k_{\rm F} \ell \gg
1$, but it is also the standard theory used to describe Coulomb
scatterers in 2D systems~\cite{kn:ando1982,kn:dassarma2003}.  For
clean graphene samples, $k_{\rm F} \ell \gtrsim 4$ at low density and
$k_{\rm F} \ell \sim 100$ at high density, making the Boltzmann theory
valid.  Sixth, the average carrier density is obtained from the
potential fluctuations using a local density (also called
Thomas-Fermi) approximation as is normally done in 2D
systems~\cite{kn:efros1993,kn:ilani2000}.  This is valid so long as
$n^* / n_{\rm imp} \gg 0.01$, which guarantees that the spacing
between electrons is much less than the length of the conducting
cluster~\cite{kn:stauffer2001}.  This condition holds both empirically
and is established {\it a posteriori} to be valid throughout the
Boltzmann transport regime.  Similarly, this demonstrates that in
current experiments the Dirac cone is always filled with electrons or
holes whose average density ranges from $0.2 \ n_{\rm imp}$ to $0.5 \
n_{\rm imp}$.  Formally, one can consider the limit $n_{\rm imp}
\rightarrow \infty$, and find that $n^* = (n_{\rm imp}/32 d^2\pi
)^{1/2}$, with no lower bound on the minimum conductivity $\sigma_0
\sim 1/\sqrt{n_{\rm imp}}$, but this ``{\it mean-field}''
approximation breaks down when the inhomogeneities become so large
that calculating the conductivity through the Boltzmann transport of
the average density $n^*$ becomes meaningless.  Seventh, we ignore
Anderson localization.  The observed experimental absence of Anderson
localization is not unique to graphene.  In 2D semiconductor
systems~\cite{kn:dassarma2005b, kn:allison2006, kn:tracy2006}, a
percolation metal-insulator transition is seen at low carrier
densities instead of a localization transition, whereas in graphene
the percolation transition is an electron-metal to hole-metal
transition without any intervening insulating phase.  Moreover, recent
numerical work~\cite{kn:bardarson2007, kn:nomura2007c} suggests that
there may be not be a metal-insulator transition in graphene.  
Naturally, one could relax
any of the above assumptions in a future calculation, but for the
purpose of comparison with current graphene experiments, these
approximations are well justified and provide a consistent framework
to understand graphene transport.  We observe that in our
RPA-Boltzmann theory, $\sigma_0$ reaches the so-called universal Dirac
point minimum conductivity
value~\cite{kn:fradkin1986,kn:ludwig1994,kn:katsnelson2006} of
$e^2/(\pi \hbar)$ for unphysically large impurity densities of $n_{\rm
imp} \gtrsim 10^{14} \rm{cm}^{-2}$, where the diffusive transport
approximation no longer applies.

The theoretical picture presented here is shown heuristically in
Fig.~\ref{Fig:model}.  Charged impurities either in the substrate or
in the vicinity of graphene create a spatially inhomogeneous potential
distribution in the graphene plane.  At low carrier density, the
spatially inhomogeneous potential breaks the system up into puddles of
electrons and holes.  This theoretical prediction~\cite{kn:hwang2006c}
has now been verified in a recent surface probe
experiment~\cite{kn:martin2007} using a scanning
single-electron-transistor to directly measure the potential
fluctuations in graphene, and finds quantitative agreement with
earlier predictions~\cite{kn:galitski2007} for the height and width of
the electron and hole puddles.  In addition, there is recent indirect
experimental support~\cite{kn:huard2007,kn:williams2007} for the
electron-hole puddle picture proposed in Ref.~\cite{kn:hwang2006c}.
Unlike usual 2D systems, both electrons and holes screen the external
potential.  These potential fluctuations directly change the local
chemical potential inducing a residual density which in turn changes
the screening.  Here we use a self-consistent procedure to determine
the residual density $n^*$, which manifests itself in experiments by a
residual conductivity plateau that is shifted by an offset gate
voltage $V_g^D = \bar{n}/\alpha$, whose width is $n^*$ and whose
magnitude is $\sigma(n^*)$, where $\sigma(n)$ is the RPA-Boltzmann
conductivity for carrier density $n$.  $V_g^D$ is called the Dirac
gate voltage because it is the value of the external gate voltage
where the Hall coefficient changes sign indicating that carriers
change from electrons to holes.  In MOSFETs, the voltage corresponding
to $V_g^D$ is often referred to as the ``{\it threshold voltage}'',
since in these systems it marks the onset of conductivity at this
critical value of carrier density, and $\sigma = 0$ below this
threshold.  In graphene, $V_g^D$ separates the conducting electron and
conducting hole transport regimes, whereas in MOSFETs the threshold
voltage separates conducting and insulating 2D channels.  Here $\alpha
\sim 7.2 ~\times 10^{10} \rm{cm}^{-2} V^{-1}$ is a geometry related
factor (that can be measured
directly~\cite{kn:novoselov2005,kn:zhang2005} using Hall measurements)
and is used to convert the experimentally measured gate voltage to
carrier density $n$.  For the rest of the paper we develop the theory
only for carrier density $n = k_{\rm F}^2/\pi$, where it is understood
that any comparison with the experimentally measured gate voltages is
made using $ n = \alpha V_g$.

\begin{figure}[t]
\vspace{0.05in}
\begin{center}
\centerline{\includegraphics[width=.4\textwidth]{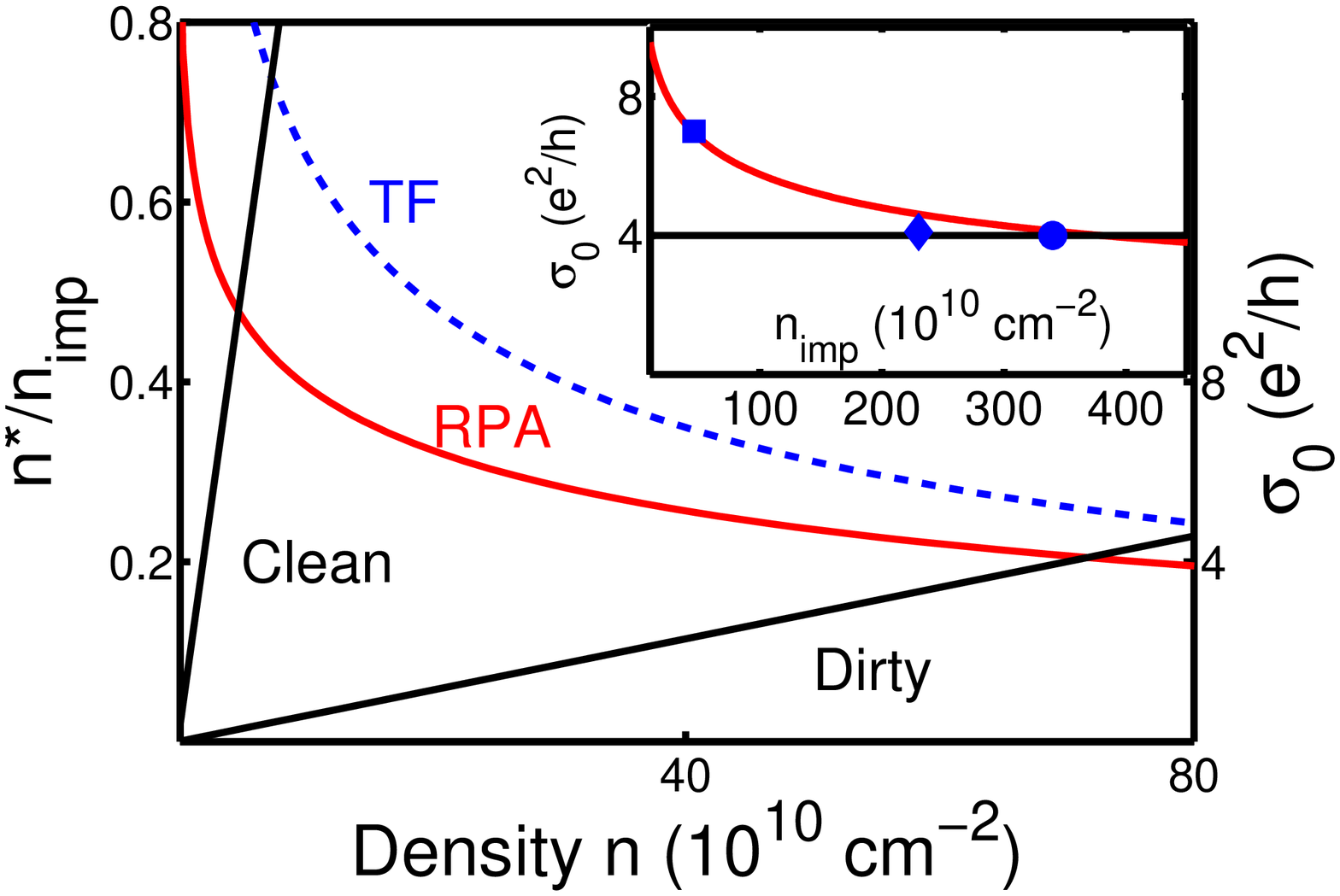}}
 \caption{\label{Fig:SCres} 
The main panel shows Eq.~\protect{\ref{Eq:SC}}.  The two lines
starting at the origin show $n/n_{\rm imp}$ for $n_{\rm imp} =
10~\times 10^{10}{\rm cm}^{-2}$ (very clean) and $n_{\rm imp} =
350~\times 10^{10}{\rm cm}^{-2}$ (very dirty).  The RPA and
Thomas-Fermi results show $2 r_s^2 C_0(r_s, a=4d \sqrt{\pi n})$ as a
function of carrier density $n$.  The points of intersection represent
the self-consistent solution, and one can read off $n^*$ from the
x-axis, $n^*/n_{\rm imp}$ from the left-axis and the \protect{``{\it
minimum conductivity}''} $\sigma_0$ from the right axis.  The inset
shows $\sigma_0$ as a function of charged impurity concentration
showing that (i) it is non-universal, (ii) dirty-samples have
$\sigma_0 = 4 e^2/h$ over a wide range of impurity concentration and
(iii) the cleanest samples have $\sigma_0 \approx 8 e^2/h$, but the
value is sensitive to the concentration of charged impurities.  Also
shown is comparison with representative experimental results, where
the square shows results from Columbia, diamond from Manchester, 
and circle from Maryland.  These same three samples (with conductivity 
shown over the full density range) were compared to a 
high density numerical Boltzmann theory in 
Ref.~\protect{\cite{kn:hwang2006c}}.}
\end{center}
\end{figure}

Based on estimates from surface probe
measurements~\cite{kn:novoselov2004,kn:ishigami2007}, and consistent
with earlier work both in
graphene~\cite{kn:hwang2006c,kn:galitski2007} and in
Si-MOSFETs~\cite{kn:ando1982,kn:dassarma2004}, we assume that the
charged impurities lie in a plane at a distance $d \sim 1 \ {\rm nm}$
from the graphene sheet and calculate the voltage fluctuations taking
into account of screening using the RPA approximation.  The screened
voltage fluctuation is a function of $d$ and the carrier density $n$;
a larger carrier density more effectively screens the charged
impurities, while the potential fluctuations are larger for low
carrier density. We include this effect self-consistently in our
theory, where $n$ is both determined by and determines the screened
impurity potential.  Our theoretical results do not depend in any
qualitative manner on the precise choice of $d$, and one can develop
relationships between the four experimental quantities (i.e. mobility,
plateau width, minimum conductivity and shift in gate voltage) that
are independent of $d$.  We emphasize that since a single parameter
$n_{\rm imp}$ determines all four experimental quantities, we
anticipate that our theory would be consistent with each of them to
within a factor of 2.  Comparison with representative
samples from the Columbia, Manchester, and Maryland groups 
(see inset of Fig.~\ref{Fig:SCres} and note added at the end of this paper)
shows agreement for the mobility and gate-voltage shift
and agrees to within the expected factor of 2 with measurements of the 
plateau width and minimum conductivity. 
   
\begin{figure}[b]
\vspace{0.05in}
\begin{center}
\centerline{\includegraphics[width=.4\textwidth]{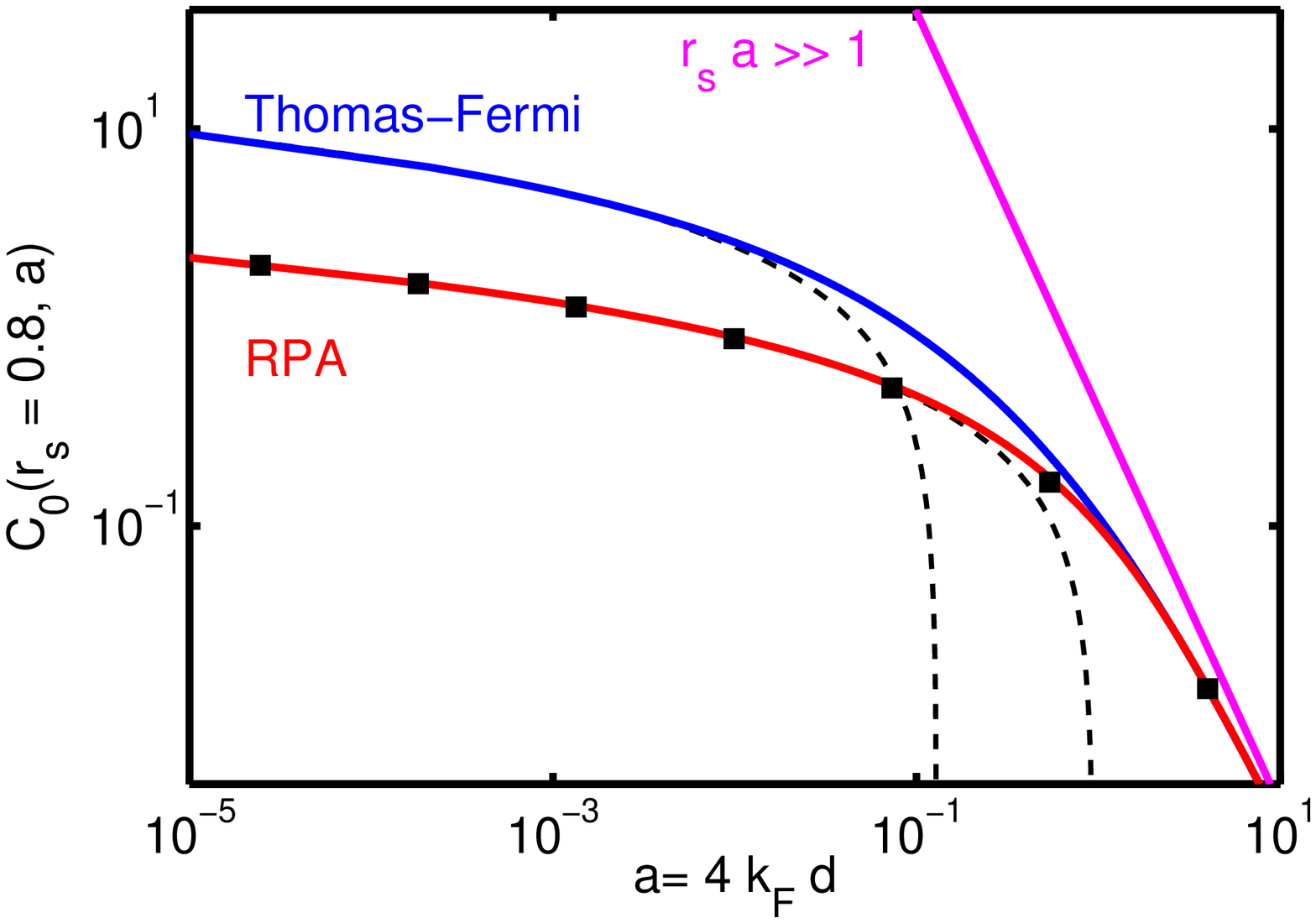}}
\caption{\label{Fig:results} 
Comparison of voltage fluctuation $C_0(r_s,a)$ using
different screening approximations.  
The Random-Phase-Approximation (RPA) is the main 
approximation used in the present work. 
The Thomas-Fermi result was derived in
Ref.~\protect{\cite{kn:galitski2007}} and the ``{\it complete screening}''
result valid for $r_s a \gg 1$, was obtained by
Ref.~\protect{\cite{kn:efros1993}}.  All three approximations agree in
the large density limit, but disagree for small density.  Shown in
dashed lines are small density analytic asymptotes for the
Thomas-Fermi and RPA and squares show the numerical evaluation
of Eq.~\protect{\ref{Eq:CORPA}} using the exact dielectric function reported in
Ref.~\protect{\cite{kn:hwang2006b}}.}
\end{center}
\end{figure}

We now proceed to calculate the Boltzmann transport conductivity.  For 2D
graphene, the semi-classical diffusive conductivity is given by
\begin{equation}
\sigma = \frac{g_s g_v e^2}{h}\frac{E_F \tau}{2 \hbar} =  
         \frac{g_s g_v e^2}{h}\frac{k_F \ell}{2 \hbar},
\end{equation}
where $g_s = g_v = 2$ are the spin and valley degeneracy 
factors and the mean free path $\ell = v_F \tau$, with 
the scattering time $\tau$ being given at $T=0$ by 
\begin{equation}
\frac{\hbar}{\tau(\vk)} = 
\frac{n_\mathrm{imp}}{4 \pi} \int d \vk' \left[\frac{V(|\vk - \vk'|)}
{\epsilon(|\vk - \vk'|)}\right]^2
[1-\cos^2(\theta)] \delta(E_{k'} - E_{k}),
\label{Eq:TI}
\end{equation}
where $V(q)=2 \pi e^{-qd} e^2/(\kappa q)$ is the Fourier Transform of
bare Coulomb potential at the transfer momentum $q=|\vk - \vk'| = 2
k_F \sin(\theta/2)$.  While the exact RPA dielectric function
is known~\cite{kn:hwang2006b}, for our purposes we can use the following
simple approximate expression which allows for analytic calculations
and provides results that are indistinguishable from the exact 
results (see Fig~\ref{Fig:results})
\begin{eqnarray}
\label{Eq:RPA}
\epsilon(q) = \left\{ \begin{array}{l}
  1 + q_s/q \ \ \ \ \mbox{if $q < 2 k_{\rm F}$}, \\
  1 + \pi r_s/2 \ \ \ \mbox{if $q > 2 k_{\rm F}$},
\end{array} \right.
\end{eqnarray}
where $q_s = 4 k_{\rm F} r_s$, and solving the integrals exactly we find
\begin{eqnarray}
\label{Eq:GX}
\sigma &=& \frac{e^2}{h}\frac{n}{n_{\rm imp}}\frac{2}{G[2 r_s]} , \\
\frac{G[x]}{x^2}  &=& \frac{\pi}{4} + 3 x - \frac{3 \pi x^2}{2} 
+ \frac{x (3 x^2 -2) \arccos~[1/x]}{\sqrt{x^2 -1}}, \nonumber
\end{eqnarray}
where for graphene on a SiO$_2$ substrate, $r_s = e^2/(\hbar \gamma
\kappa) \approx 0.8$, $G(2r_s) \approx 1/10$ and $ \sigma \approx 20
(e^2/h) (n/n_{\rm imp})$.  Note that $G[x]$ is positive and real 
for all $x$.   From this we derive a simple analytical
expression for linear tail mobility in graphene
\begin{equation}
\label{Eq:mobility}
\frac{\mu}{\mu_0} \approx 50 \frac{n_0}{n_{\rm imp}},
\end{equation} where $\mu_0 = 1 ~m^2 /Vs$ and 
$n_0 = 10^{10} \rm{cm}^{-2}$.  Note that Eq.~\ref{Eq:mobility} depends
only on the charged impurity scattering concentration $n_{imp}$,
indicating that the only way to increase graphene mobility for fixed
$r_s$ is to improve the sample quality.  However, Eq.~\ref{Eq:GX} also
shows that mobility depends on the substrate dielectric constant
$\kappa$, and therefore changing the underlying substrate from SiO$_2$
to one with a higher dielectric constant would reduce $r_s$, and would
be another way to increase sample mobility.  For example, 
changing the substrate to Hf0$_2$ ($\kappa_s \sim 25$) from
SiO$_2$ ($\kappa_s \sim 4$), should enhance graphene mobility
by a factor of 5. 

We now calculate the statistics of the random voltage fluctuations
significantly extending earlier numerical work~\cite{kn:efros1993} 
to incorporate analytically the non-linear screening of electrons and 
holes in a zero-gap situation, to find  
\begin{eqnarray}
\overline{\delta V^2} &=& \overline{[V-V_g^D]^2} =
n_{\rm imp} \int \frac{d^2 q}{(2 \pi)^2}
\left[ \frac{2 \pi e^2 e^{-q d}}{\kappa q \epsilon(q)} \right]^2,
\label{Eq:CORPA1}
\\
&=& 2 \pi n_{\rm imp} \left(\frac{e^2}{\kappa}\right)^2 C_0 (r_s, a= 4 k_{\rm F} d), \nonumber \\
C_0^{\rm RPA}(r_s, a) &=& -1 + \frac{4 E_1(a)}{(2 + \pi r_s)^2}
+ \frac{2 e^{-a} r_s }{1 + 2 r_s} \label{Eq:CORPA} \\
&& \mbox{} + (1 + 2 r_s a)~e^{2 r_s a} (E_1[2 r_s a] - E_1[a (1+ 2 r_s)]),
\nonumber 
\end{eqnarray}
where the superscript on $C_0^{\rm RPA}$ indicates that we used the
RPA Approximation, and $E_1(z)= \int_z^\infty t^{-1} e^{-t} dt$ is the 
exponential integral function.  The voltage fluctuation 
result $C_0(r_s = 0.8, a)$ is shown in Fig.~\ref{Fig:results} 
comparing different approximation schemes used in the literature.  
The analytic result Eq.~\ref{Eq:CORPA} is compared with a numerical 
evaluation of Eq.~\ref{Eq:CORPA1} using the exact
dielectric function first reported in Ref.~\cite{kn:hwang2006b}, 
as well as with the long-wavelength (also known as Thomas-Fermi)
approximation~\cite{kn:galitski2007}, where $\epsilon(q) = 1 + q_s/q$ 
for all $q$, and the ``{\it complete screening}''
approximation~\cite{kn:nomura2007,kn:efros1993} that is valid only for
$r_s a \gg 1$, where $\epsilon(q) = q_s/q$.  In the limit of 
large $a$ (i.e. $k_{\rm F} d \gg 1$), both the Thomas-Fermi 
and the RPA result approach the complete screening limit of 
$C_0^{\rm CS}(r_s, a) = (2 r_s a)^{-2}$, that was previously
obtained in Ref.~\cite{kn:efros1993}. The three approximations disagree 
in the small $a$ (or low density) limit where
\begin{equation}
C_0^{\rm RPA}(r_s, a \rightarrow 0) = 
\frac{-1}{2 r_s + 1} - \ln \left[\frac{2 r_s}{2 r_s +1}\right]
- \frac{4 \ln(\tilde{\gamma} a)}{(2 + \pi r_s)^2},
\end{equation}     
and $\tilde{\gamma} \approx 1.781$ is Euler's constant.  Notice that for 
RPA, a combination of the small and large $a=4 k_{\rm F} d$ asymptotes span
most of the density range. While these analytical asymptotes
correctly describe the screened potential for most densities, it turns
out, that the regime relevant to graphene experiments is the window
where they do not work well, and the full functional form of $C_0^{\rm
RPA}$ shown in Eq.~\ref{Eq:CORPA} needs to be used.

As discussed earlier, to determine the
self-consistent residual density, we equate the average chemical
potential to the fluctuation in the screened charged impurity induced
potential as $\overline{E_{\rm F}^2} = \overline{\delta V^2}$, and
find
\begin{eqnarray}
\label{Eq:SC}
\frac{n^*}{n_{\rm imp}} &=& 2 r_s^2 C_0^{\rm RPA}(r_s, a = 4 d \sqrt{\pi
n^*}), \nonumber \\
\bar{n} &=& \frac{n_{\rm imp}^2}{4 n^*},
\end{eqnarray}
where the second expression for the impurity induced shift in
voltage is determined from $\overline{V} = \pi n_{\rm imp}
\gamma/(2 k_{\rm F})$~\cite{kn:efros1993,kn:galitski2007}.  
Combining these results gives Eq~\ref{Eq:Main}.

Shown in the main panel of Fig.~\ref{Fig:SCres} are the results of the
self-consistent procedure.  The inset shows the value of the minimum
conductivity and compares with the experimental results.  These are the
same three samples that were shown in Ref.~\cite{kn:hwang2006c} to
compare the conductivity at {\it high} carrier density (far from the Dirac
point) with a numerical Boltzmann theory, and here we show 
the {\it low} carrier density comparison near the Dirac point.  Through
Eq.~\ref{Eq:mobility}, we have the high-density measurements directly
giving $n_{\rm imp}$, and this is the only parameter used to determine
the minimum conductivity $\sigma_0$.  Our results show that contrary
to common perception, the graphene minimum conductivity is not
universal, but that future cleaner samples will have higher values of
$\sigma_0$.  We emphasize that, in addition to explaining the value of
$\sigma_0$ and its dependence on the sample quality, our theory also
naturally accounts for the width of the minimum conductivity plateau
in agreement with
experiments.  For
example, $n_{\rm imp} = 350 \times 10^{10}{\rm cm}^{-2}$ gives
$n^{*} = 70 \times 10^{10}{\rm cm}^{-2}$ and plateau width $\Delta V_g
= 10~V$.

We emphasize that the most important qualitative result of our theory
is to introduce a realistic mechanism operational in all disordered
graphene samples (i.e. in the presence of random charged impurities)
which produce a plateau-like approximate non-universal minimum
graphene conductivity at low induced carrier density.  We obviously
can not rule out other possible ``{\it universal mechanisms}'' which
will lead to a ``{\it universal}'' minimum intrinsic graphene
conductivity at the Dirac point in the clean limit, a situation beyond
the scope of our theory.  But the fact that the currently existing
experimental data from three different groups exhibit non-universal
minimum conductivity in approximate (within a factor of 2) agreement
with our theory indicates that any intrinsic universal mechanism
beyond our model may not yet be playing any role.  We note however, 
that there have been recent theoretical
~\cite{kn:katsnelson2006,kn:tworzydlo2006,kn:fertig2006} and
experimental ~\cite{kn:miao2007} work dealing with ballistic transport
in mesoscopic graphene which show universal behavior in the regime
where $\ell> W > L$, where $\ell$ is the mean free path, and $W$ and
$L$ are the sample width and length.  Our theory does not apply in
this zero disorder ballistic limit since our work is entirely 
built on the picture of diffusive transport through disorder 
induced electron-hole puddles.

In summary, we believe we have qualitatively and semi-quantitatively
solved one of the main transport puzzles in graphene, namely, why the
experimentalists see a conductivity minimum plateau and the extent to
which this minimum conductivity is or is not universal.  The theory
developed here should only be taken as the first step toward a full
quantitative theory of graphene transport, particularly at the lowest
carrier densities.  Many questions still remain open, although we
believe that we have taken an important step in the right direction.
In particular, the precise nature of transport at the charge neutral
Dirac point can not be accessed by our self-consistent treatment which
is valid only at finite doping away from charge neutrality.  The
percolation through electron and hole puddles becomes precisely
equivalent at the Dirac point, and a purely percolative theory, not
our self-consistent RPA-Boltzmann theory, would be necessary to
understand the transport.  We do, however, mention that the observed
smooth behavior of conductivity as a function of gate voltage through
the charge neutrality point indicates a lack of any dramatic phenomena
at the Dirac point, and given that our theory is a good description of
transport away from the Dirac point, it is conceivable that it remains
quantitatively valid at the charge neutral point also.  Given the great 
deal of current interest in graphene and the potential
for graphene-based electronics applications, our transport theory of
graphene not only furthers our understanding of this new material, but
it provides essential insights on how to obtain higher mobility, which
is necessary if graphene is to have serious technological impact 
as a new electronic material.

\begin{acknowledgments}
This work is supported by the US-ONR, LPS-NSA, NSF-NRI and the 
Microsoft Project Q.  It is a pleasure to thank M. Fuhrer, 
A. Geim, P. Kim and H. Stormer for sharing with us their
experimental data. {\it Note Added:} After submission of this 
manuscript, there have been two experimental studies 
by Ref.~\cite{kn:tan2007} and Ref.~\cite{kn:chen2007b} who find 
that our transport
theory is in good agreement with experimental data taken over a wide range
of charged impurity densities.
\end{acknowledgments}


\end{article}
\end{document}